\documentclass[twocolumn,floats,floatfix,aps,pra]{revtex4}

\usepackage{amsfonts}
\usepackage{amssymb}
\usepackage{amsmath}
\usepackage{calc}
\usepackage[dvips]{graphicx}
\usepackage{bm}
\usepackage{mathrsfs}

\def\be{\begin{equation}}
\def\ee{\end{equation}}
\def\bea{\begin{eqnarray}}
\def\eea{\end{eqnarray}}
\def\bse{\begin{subequations}}
\def\ese{\end{subequations}}

\def\1{\mathbf{1}}

\begin{document}

\author{Lachezar S. Simeonov}
\affiliation{Department of Physics, Sofia University, James Bourchier 5 blvd, 1164 Sofia, Bulgaria}

\title{Intuitive Derivation of the Coriolis Force}
\date{\today }

\begin{abstract}
The major difficulty when one teaches about non-inertial reference frames in undergraduate courses on Classical Mechanics is to find an \textit{intuitive} way to derive the Coriolis acceleration. Indeed, there is a factor of 2 in the formula for the Coriolis acceleration and this factor is shrouded in mystery. In this paper we not only show an intuitive way to \textit{derive} the Coriolis acceleration but we also show \textit{why} there is a factor of 2. Indeed, it turns out that the Coriolis acceleration results from \textit{two} completely different reasons (and hence the factor of 2). The \textit{first} reason is this - as the particle moves to a new position, it `sees` a different local velocity of the rotating frame. The \textit{second} reason is purely geometrical - the velocity vector is subjected to purely geometrical rotation due to the rotation of the reference frame. \textit{Both} of these contributions unite and they result in the Coriolis acceleration.
\end{abstract}


\maketitle

\section{Introduction}
In the standard undergraduate (as well as graduate) courses on Classical Mechanics \cite{Sommerfeld66, Landau2000, Goldstein1980, Thornton2004, Greiner2003} people usually derive the motion of bodies in non-inertial reference frames in a rigorous but not very intuitive way. Indeed, as a custom, authors of textbooks derive the Coriolis and centrifugal forces by merely taking derivatives of rotating vectors. In this manner they derive the familiar Coriolis acceleration: $\textbf{a}_{C}=-2\boldsymbol{\Omega}\times \textbf{v}$. Here $\textbf{v}$ is the velocity vector of a particle as seen in the rotating reference frame and $\boldsymbol{\Omega}$ is the angular velocity of the rotating frame itself. We see that there is a factor of $2$ in the Coriolis acceleration. As far as the author of this work knows this factor is not explained intuitively anywhere. Indeed, Persson \cite{Persson1998} derives the Coriolis force from dynamical but not kinematical principles. It is typical for books on meteorology to derive the Coriolis force from the law of conservation of angular momentum \cite{Holton2004, Byers1959, Silver2011}, which however does not explain the actual \textit{kinematic} nature of the Coriolis force and the factor of 2 is still mysterious. Herrera \textit{et al.} \cite{Herrera2016} give another dynamical derivation of the Coriolis force. We can find detailed historical account of the Coriolis force in Ref. \cite{Graney2011, Graney2016, Dugas1988} and yet nowhere we see a satisfactory and intuitive \textit{kinematic} approach to the Coriolis acceleration.

In this paper we derive \textit{intuitively} the expression for the Coriolis force. We show why there is a factor of 2 in the Coriolis acceleration: $\textbf{a}_{C}=-2\boldsymbol{\Omega}\times \textbf{v}$. It turns out that the Coriolis acceleration results from \textit{two} completely different reasons (and this explains the factor of 2). The \textit{first} reason is this - as the particle moves to a new position, it `sees` a different local velocity of the rotating frame. The \textit{second} reason is purely geometrical - the velocity vector is subjected to purely geometrical rotation due to the rotation of the reference frame. Both of these effects have separate contributions to the final result, which is the Coriolis acceleration.

The paper is organized as follows: In Section II we start with the notation that will be used throughout this work. In Section III we derive the general formula for rotating a vector along an arbitrary axis, which will be important later. In Section IV we derive the Coriolis and the centrifugal acceleration in the standard way, which is quite rigorous but not very intuitive. In Section V we consider special scenarios of motion of a body examined in a rotating frame. We do this in order to test our formulas for the Coriolis and the centrifugal accelerations and to supply intuition that will be necessary later. In Section VI we finally derive intuitively the Coriolis acceleration. In Section VII we give conclusions and summary.

\section{Notation}
We denote with $XYZ$ the \textit{inertial} (which we shall also call \textit{stationary}) reference frame. The vectors $\textbf{I}, \textbf{J}$ and $\textbf{K}$ are the unit basis vectors in the $X, Y$ and $Z$ axes respectively. We denote with the small letters $xyz$ the \textit{rotating} reference frame, while $\textbf{i}, \textbf{j}$ and $\textbf{k}$ are the \textit{rotating} unit basis vectors of this frame (see FIG. 1). 

\begin{figure}[tb]
\includegraphics[width= 1.0\columnwidth]{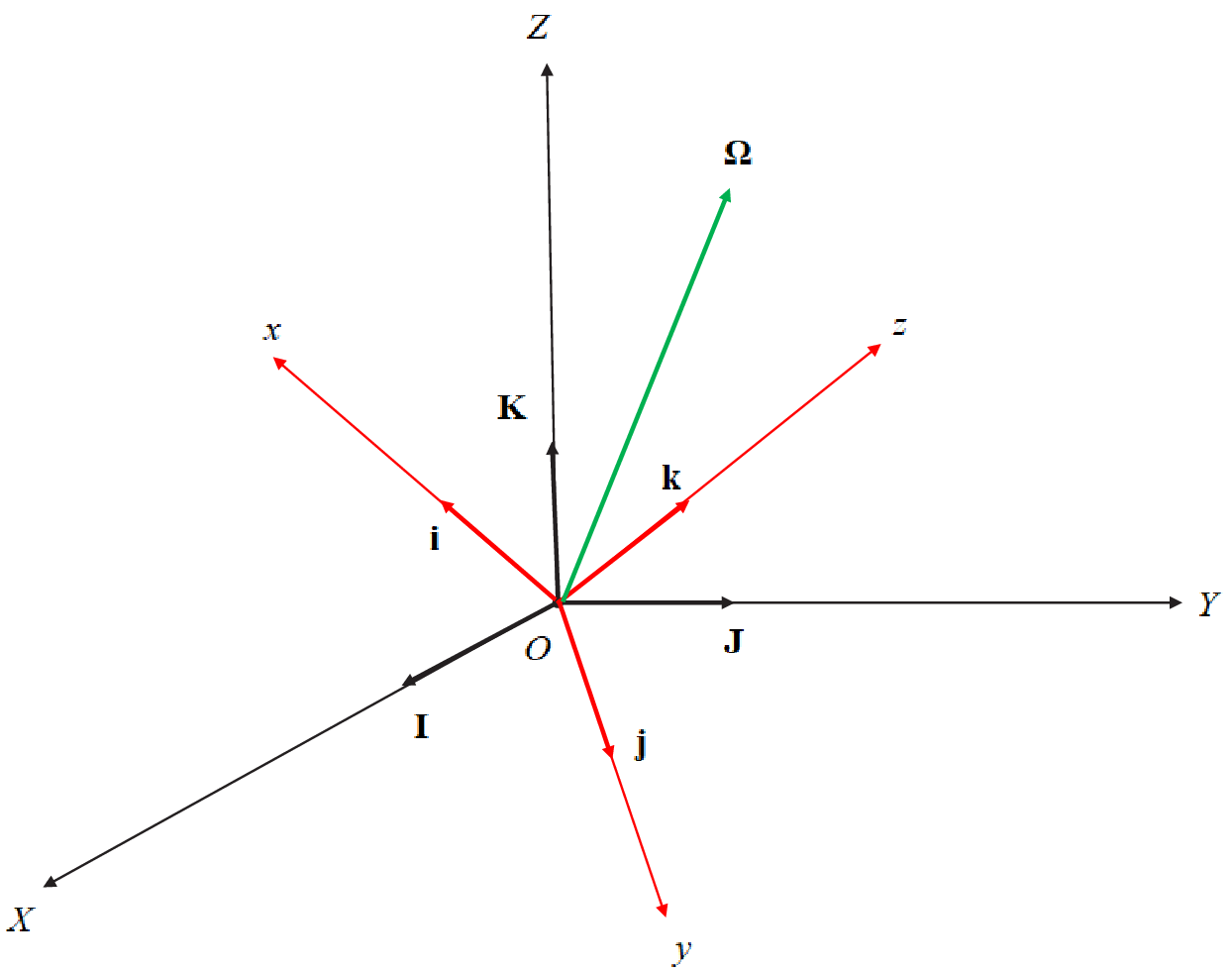}
\caption{(Color online) Inertial frame $XYZ$ and a rotating frame $xyz$ with a common origin $O$. The stationary unit basis vectors are called $\textbf{I}$, $\textbf{J}$ and $\textbf{K}$, while the rotating unit vectors are called $\textbf{i}$, $\textbf{j}$ and $\textbf{k}$.}
\end{figure}

Obviously, the stationary observer sees that $\textbf{i}, \textbf{j}$ and $\textbf{k}$ change in time: $\textbf{i}=\textbf{i}\left(t\right)$, $\textbf{j}=\textbf{j}\left(t\right)$ and $\textbf{k}=\textbf{k}\left(t\right)$.

Both coordinate systems have common origin - the point $O$. We denote with $\boldsymbol{\Omega}$ the angular velocity of the rotating frame. In this paper we consider, for  simplicity's sake the case when $\boldsymbol{\Omega}=\text{const.}$, i.e. $\boldsymbol{\Omega}$ does \textit{not} change in time.

We shall use \textit{capital} letters to denote all quantities (like position vector, velocity, acceleration) in the \textit{stationary} reference frame. Thus, the position vector is:

\begin{equation}
\textbf{R}\left(t\right)=X\left(t\right)\textbf{I}+Y\left(t\right)\textbf{J}+Z\left(t\right)\textbf{K}.
\end{equation}

The velocity vector is:
\begin{equation}
\textbf{V}\left(t\right)=\frac{d\textbf{R}}{dt}=\frac{dX}{dt}\textbf{I}+\frac{dY}{dt}\textbf{J}+\frac{dZ}{dt}\textbf{K}.
\end{equation}

The acceleration vector for the stationary observer is:
\begin{equation}
\textbf{A}\left(t\right)=\frac{d^{2}\textbf{R}}{dt^{2}}=\frac{d^{2}X}{dt^{2}}\textbf{I}+\frac{dY^{2}}{dt^{2}}\textbf{J}+\frac{d^{2}Z}{dt^{2}}\textbf{K}.
\end{equation}

We shall use \textit{small} letters to denote all quantities in the \textit{rotating} frame. Thus, the position vector is
\begin{equation}
\textbf{r}\left(t\right)=x\left(t\right)\textbf{i}\left(t\right)+y\left(t\right)\textbf{j}\left(t\right)+z\left(t\right)\textbf{k}\left(t\right).\label{rrot}
\end{equation}

Now, obviously
\begin{equation}
\textbf{R}=\textbf{r}.\label{R=r}
\end{equation}

Indeed, both vectors are in fact the \textit{same vector} because this vector points from the common origin $O$ to the position of the particle. It is just that the \textit{coordinates} of this vector $\textbf{r}=\textbf{R}$ are different for the different reference frames. For this reason we can interchange the vector $\textbf{R}$ with $\textbf{r}$ and vice versa.

We use the small letter $\textbf{v}$ to denote the velocity of the particle in the \textit{rotating} frame. Note that:
\begin{equation}
\textbf{v}\neq\frac{d\textbf{r}}{dt}.
\end{equation} 
Indeed,
\begin{equation}
\frac{d\textbf{r}}{dt}=\frac{dx}{dt}\textbf{i}+\frac{dy}{dt}\textbf{j}+\frac{dz}{dt}\textbf{k}+x\frac{d\textbf{i}}{dt}+y\frac{d\textbf{j}}{dt}+z\frac{d\textbf{k}}{dt}.\label{drdt}
\end{equation}
In other words, \textit{we have to take the derivatives of the basis vectors}.  However the rotating observer \textit{does not see that the basis vectors rotate}. Indeed, from their point of view $\textbf{i},\textbf{j},\textbf{k}=\text{const.}$ Of course, $\textbf{i},\textbf{j},\textbf{k}\neq\text{const.}$ but the rotating observer does \textit{not} see this fact. For that reason,
\begin{equation}
\textbf{v}=\frac{dx}{dt}\textbf{i}+\frac{dy}{dt}\textbf{j}+\frac{dz}{dt}\textbf{k}\neq\frac{d\textbf{r}}{dt}.\label{vrot}
\end{equation}

Similarly, the acceleration $\textbf{a}$ for the rotating observer is:
\begin{equation}
\textbf{a}=\frac{d^{2}x}{dt^{2}}\textbf{i}+\frac{d^{2}y}{dt^{2}}\textbf{j}+\frac{d^{2}z}{dt^{2}}\textbf{k}\neq\frac{d^{2}\textbf{r}}{dt^{2}}.\label{arot}
\end{equation}
Sometimes people invent different notations for the derivatives in the inertial and the non-inertial reference frames. We shall avoid such notations.

\section{How to rotate a vector}
From Eq. \eqref{drdt} we see that we need to find the derivatives $d\textbf{i}/dt$, $d\textbf{j}/dt$, and $d\textbf{k}/dt$ for the rotating vectors $\textbf{i}$, $\textbf{j}$ and $\textbf{k}$.

To that end let us explore how some \textit{arbitrary} vector $\textbf{B}$ rotates by a small angle $\delta\varphi$ along some arbitrary axis. The reader can consult FIG. 2. 
\begin{figure}[tb]
\includegraphics[width= 1.0\columnwidth]{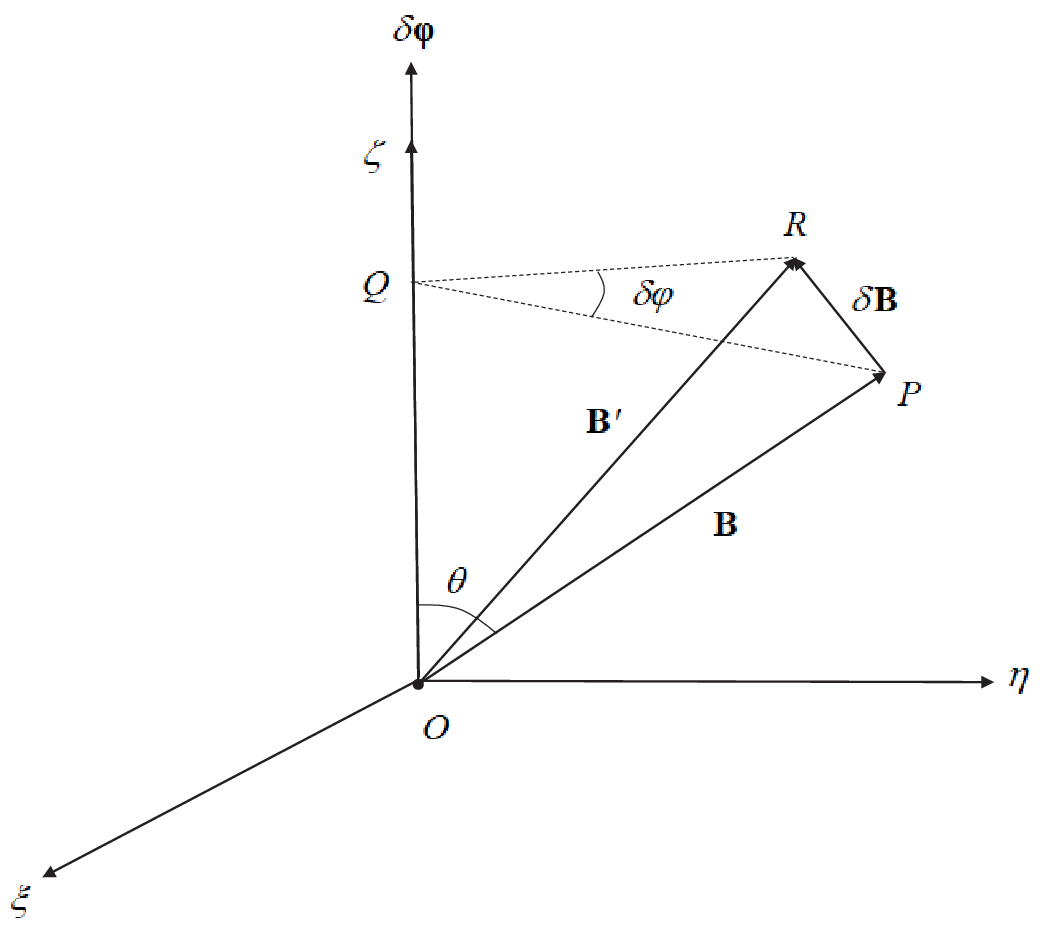}
\caption{Rotation of an arbitrary vector $\textbf{B}$ to a new vector $\textbf{B}^{\prime}$. Here $\delta\textbf{B}=\textbf{B}^{\prime}-\textbf{B}$. The $\zeta$ axis is the axis of rotation and $\delta\varphi$ is the angle of rotation. The vector $\delta\boldsymbol{\varphi}$ points in the axis of rotation and $|\delta\boldsymbol{\varphi}|=\delta\varphi$.}
\end{figure}

Let us invent a vector $\delta\boldsymbol{\varphi}$ to be such that it has the same direction as the axis of rotation (see FIG. 2) and whose magnitude is 
\begin{equation}
|\delta\boldsymbol{\varphi}|=\delta\varphi.
\end{equation}

We shall prove the following formula:
\begin{equation}
\delta\textbf{B}\approx\delta\boldsymbol{\varphi}\times \textbf{B},\label{rot}
\end{equation}
which is true if we ignore higher order quantities $\delta\varphi^{2},\delta\varphi^{3},...$. Here $\delta\textbf{B}$ is the difference between the rotated vector $\textbf{B}^{\prime}$ and initial vector $\textbf{B}$ (FIG. 2). In other words,
\begin{equation}
\delta \textbf{B}=\textbf{B}^{\prime}-\textbf{B}.
\end{equation} 

First, let us erect a coordinate system $\xi,\eta,\zeta$. And let the $\zeta$ axis points in the same direction as the axis of rotation, i.e. in the same direction as the vector $\delta\boldsymbol{\varphi}$. Then the rotation of the vector $\textbf{B}$ is expressed through the standard rotation matrix \cite{Goldstein1980}:
\begin{equation}
\left(\begin{array}{c}
B_{\xi} ^{\prime}\\
B_{\eta}^{\prime}\\
B_{\zeta}^{\prime} 
\end{array}\right)=\left(\begin{array}{ccc}
\cos{\delta\varphi} & -\sin{\delta\varphi} & 0\\
\sin{\delta\varphi} & \cos{\delta\varphi} & 0\\
0 & 0 & 1
\end{array}\right).\left(\begin{array}{c}
B_{\xi}\\
B_{\eta}\\
B_{\zeta} 
\end{array}\right).
\end{equation}

We replace $\cos{\delta\varphi}\approx 1$ and $\sin{\delta\varphi}\approx\delta\varphi$ and we have:

\begin{equation}
\left(\begin{array}{c}
B_{\xi} ^{\prime}\\
B_{\eta}^{\prime}\\
B_{\zeta}^{\prime} 
\end{array}\right)\approx\left(\begin{array}{ccc}
1 & 0 & 0\\
0 & 1 & 0\\
0 & 0 & 1
\end{array}\right).\left(\begin{array}{c}
B_{\xi}\\
B_{\eta}\\
B_{\zeta} 
\end{array}\right)+\left(\begin{array}{ccc}
0 & -\delta\varphi & 0\\
\delta\varphi & 1 & 0\\
0 & 0 & 1
\end{array}\right).\left(\begin{array}{c}
B_{\xi}\\
B_{\eta}\\
B_{\zeta} 
\end{array}\right)\label{matrix1}
\end{equation}
Since the vector $\delta\boldsymbol{\varphi}$ is in the $\zeta$ direction, the vector $\delta\boldsymbol{\varphi}=(\delta\varphi_{\xi},\delta\varphi_{\eta},\delta\varphi_{\zeta})=(0,0,\delta\varphi)$. Then Eq. \eqref{matrix1} can be rewritten to:
\begin{equation}
\textbf{B}^{\prime}\approx\textbf{B}+\delta\boldsymbol{\varphi}\times\textbf{B},
\end{equation}
The above equation is coordinate free and therefore does not depend on our choice of the $\xi\eta\zeta$ coordinate system. In this way we have proved Eq. \eqref{rot}.

We can show Eq. \eqref{rot} in another way. Indeed, if we consider the small vector $\delta \textbf{B}$ as an arc of a small circle (see FIG. 2), then obviously, $|\delta \textbf{B}|\approx PQ.\delta\varphi$. On the other hand $PQ=|\textbf{B}|\sin{\theta}$ and we reach the conclusion that $|\delta \textbf{B}|\approx |\textbf{B}|\delta\varphi\sin{\theta}=|\delta\boldsymbol{\varphi}\times\textbf{B}|$. By the right hand rule we see that the direction  of the small vector $\delta\textbf{B}$ is the same as the direction of  the vector $\delta\boldsymbol{\varphi}\times\textbf{B}$. This shows again that Eq. \eqref{rot} is true.

If the rotation of the vector $\textbf{B}$ has happened for a small time interval $\delta t$, and if we divide Eq. \eqref{rot} by $\delta t$ we have:
\begin{equation}
\frac{\delta \textbf{B}}{\delta t}\approx\frac{\delta\boldsymbol{\varphi}}{\delta t}\times\textbf{B}.\label{rot2}
\end{equation}
We define the angular velocity:
\begin{equation}
\boldsymbol{\Omega}\equiv\lim_{\delta t\rightarrow 0}\frac{\delta\boldsymbol{\varphi}}{\delta t}.
\end{equation}
We take the limit $\delta t\rightarrow 0$ in Eq. \eqref{rot2} and the approximate equations become true equations:
\begin{equation}
\frac{d\textbf{B}}{dt}=\lim_{\delta t\rightarrow 0}\frac{\delta \textbf{B}}{\delta t}=\lim_{\delta t\rightarrow 0}\frac{\delta\boldsymbol{\varphi}}{\delta t}\times\textbf{B}=\boldsymbol{\Omega}\times\textbf{B}.
\end{equation}

\textit{Conclusion}: for any vector $\textbf{B}$, which rotates with angular velocity $\boldsymbol{\Omega}$, we  have that:
\begin{equation}
\frac{d\textbf{B}}{dt}=\boldsymbol{\Omega}\times\textbf{B}.
\end{equation}

In the special case of rotating a coordinate system, the basis vectors rotate as following:
\begin{align}
&\frac{d\textbf{i}}{dt}=\boldsymbol{\Omega}\times \textbf{i},\notag\\
&\frac{d\textbf{j}}{dt}=\boldsymbol{\Omega}\times \textbf{j},\label{rotbasis}\\
&\frac{d\textbf{k}}{dt}=\boldsymbol{\Omega}\times \textbf{k}.\notag
\end{align}

Before we perform the actual derivation of the Coriolis force, we have to consider one final point. Let us rotate a vector \textit{field} $\textbf{B}$ (see FIG. 3). 

\begin{figure}[tb]
\includegraphics[width= 0.7\columnwidth]{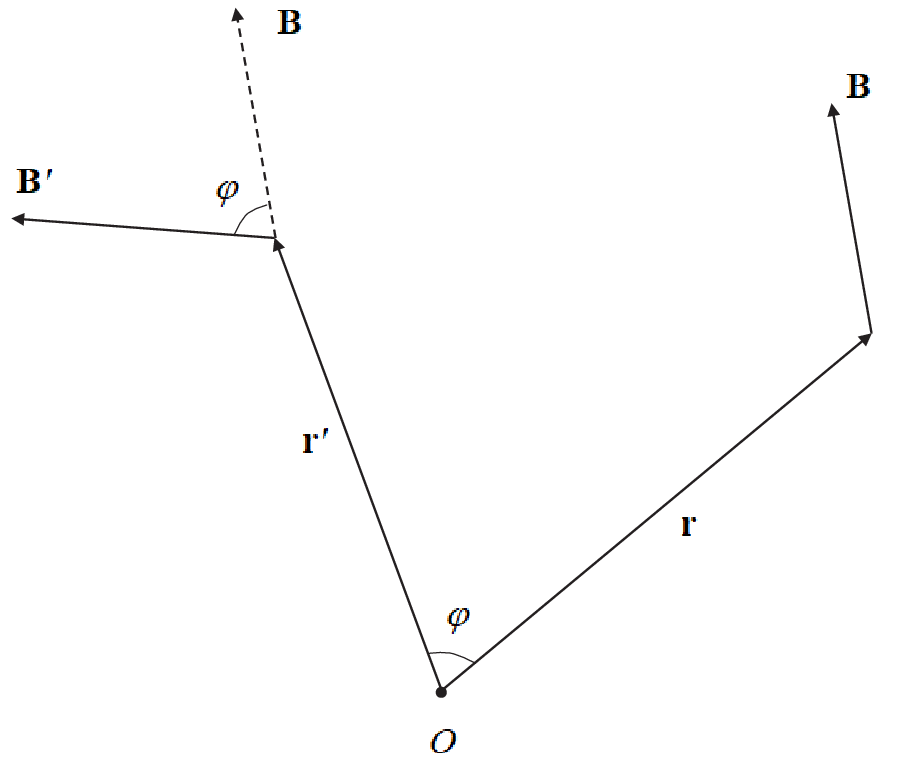}
\caption{Rotation of a vector \textit{field} $\textbf{B}$. By rotation $\textbf{r}\rightarrow \textbf{r}^{\prime}$ and then $\textbf{B}(\textbf{r})\rightarrow \textbf{B}^{\prime}(\textbf{r}^{\prime})$. The new vector $\textbf{B}^{\prime}$ is translated to the new position $\textbf{r}^{\prime}$ but is \textit{also} rotated by an angle $\varphi$. To see this rotation, we compare $\textbf{B}^{\prime}$ with the dashed vector $\textbf{B}$ and we see that the angle between $\textbf{B}$ and $\textbf{B}^{\prime}$ is again $\varphi$.}
\end{figure}

If we rotate the position vector $\textbf{r}$ by some arbitrary angle $\varphi$ (not necessarily small), then the vector field $\textbf{B}$ is translated to a new position $\textbf{r}^{\prime}$ but is \textit{also} \textit{rotated} by the \textit{same} angle $\varphi$. We can see that if we compare $\textbf{B}^{\prime}$ and $\textbf{B}$ in FIG. 3.

We are finally ready to actually derive the Coriolis acceleration.

\section{Standard Way to Derive the Coriolis and the centrifugal accelerations}

In this section we shall derive the Coriolis and the centrifugal accelerations in the standard way. We simply compare the velocities and the accelerations of a particle in the \textit{two} reference frames. This method is typical for books on Classical Mechanics \cite{Sommerfeld66, Landau2000, Goldstein1980, Thornton2004, Greiner2003}. Though it is rigorous, it is hardly intuitive. In the next sections we shall derive the Coriolis acceleration in an intuitive way.

Let us consider the motion of a particle in the two reference frames - a stationary reference frame $XYZ$ and a rotating reference frame $xyz$. We start with the obvious statement (see Eq. \eqref{R=r}):
\begin{equation}
\textbf{R}=\textbf{r}.
\end{equation}

The velocity of the particle $\textbf{V}$ according to the stationary observer is:
\begin{equation}
\textbf{V}=\frac{d\textbf{R}}{dt}=\frac{d\textbf{r}}{dt}=\frac{d}{dt}\left[ x(t)\textbf{i}(t)+y(t)\textbf{j}(t)+z(t)\textbf{k}(t)\right].
\end{equation}
We simplify,
\begin{equation}
\textbf{V}=\dot{x}\textbf{i}+\dot{y}\textbf{j}+\dot{z}\textbf{k}+x\frac{d\textbf{i}}{dt}+y\frac{d\textbf{j}}{dt}+z\frac{d\textbf{k}}{dt}.
\end{equation}

We shall use both dot and $d/dt$ as symbols for time-derivative. The first three terms give the velocity $\textbf{v}$ of the particle in the rotating frame (see Eq. \eqref{vrot}). For the last three terms we use Eqs. \eqref{rotbasis}. We obtain:
\begin{align}
\textbf{V}=\textbf{v}+x\left(\boldsymbol{\Omega}\times\textbf{i}\right)+y\left(\boldsymbol{\Omega}\times\textbf{j}\right)+z\left(\boldsymbol{\Omega}\times\textbf{k}\right).
\end{align}
We simplify,
\begin{align}
\textbf{V}=\textbf{v}+\boldsymbol{\Omega}\times\left(x\textbf{i}+y\textbf{j}+z\textbf{k}\right).
\end{align}
But we have that $\textbf{R}=\textbf{r}=x\textbf{i}+y\textbf{j}+z\textbf{k}$. Therefore we finally have,
\begin{equation}
\textbf{V}=\textbf{v}+\boldsymbol{\Omega}\times\textbf{r}.\label{Vv}
\end{equation}
We wish to calculate the accelerations. To this end we take time derivative of this equation and recall that $\boldsymbol{\Omega}=\text{const}.$
\begin{equation}
\textbf{A}=\frac{d\textbf{V}}{dt}=\frac{d\textbf{v}}{dt}+\boldsymbol{\Omega}\times\frac{d\textbf{r}}{dt}.
\end{equation}
We use Eqs. \eqref{vrot} and \eqref{Vv} and we have:
\begin{equation}
\textbf{A}=\frac{d}{dt}\left(\dot{x}\textbf{i}+\dot{y}\textbf{j}+\dot{z}\textbf{k}\right)+\boldsymbol{\Omega}\times\left(\textbf{v}+\boldsymbol{\Omega}\times\textbf{r}\right).
\end{equation}
We simplify again,
\begin{align}
\textbf{A}&=\left(\ddot{x}\textbf{i}+\ddot{y}\textbf{j}+\ddot{z}\textbf{k}\right)+\left(\dot{x}\frac{d\textbf{i}}{dt}+\dot{y}\frac{d\textbf{j}}{dt}+\dot{z}\frac{d\textbf{k}}{dt}\right)+\notag\\
&+\boldsymbol{\Omega}\times\left(\textbf{v}+\boldsymbol{\Omega}\times\textbf{r}\right).
\end{align}
The first three terms give the acceleration $\textbf{a}$ in the rotating frame (see Eq. \eqref{arot}). Then if we use Eqs. \eqref{rotbasis}, we obtain:
\begin{equation}
\textbf{A}=\textbf{a}+\dot{x}\left(\boldsymbol{\Omega}\times\textbf{i}\right)+\dot{y}\left(\boldsymbol{\Omega}\times\textbf{j}\right)+\dot{z}\left(\boldsymbol{\Omega}\times\textbf{k}\right)+\boldsymbol{\Omega}\times\left(\textbf{v}+\boldsymbol{\Omega}\times\textbf{r}\right).
\end{equation}
We simplify again:
\begin{equation}
\textbf{A}=\textbf{a}+\boldsymbol{\Omega}\times\left(\dot{x}\textbf{i}+\dot{y}\textbf{j}+\dot{z}\textbf{k}\right)+\boldsymbol{\Omega}\times \textbf{v}+\boldsymbol{\Omega}\times\left(\boldsymbol{\Omega}\times \textbf{r}\right).
\end{equation}
We use Eq. \eqref{vrot} again and we finally obtain,
\begin{equation}
\textbf{A}=\textbf{a}+2\boldsymbol{\Omega}\times\textbf{v}+\boldsymbol{\Omega}\times\left(\boldsymbol{\Omega}\times \textbf{r}\right).
\end{equation}
Let us express $\textbf{a}$:
\begin{equation}
\textbf{a}=\textbf{A}-2\boldsymbol{\Omega}\times\textbf{v}-\boldsymbol{\Omega}\times\left(\boldsymbol{\Omega}\times \textbf{r}\right).\label{afinal}
\end{equation}
The second term on the right is the Coriolis acceleration: $-2\boldsymbol{\Omega}\times\textbf{v}$, while the last term is the centrifugal acceleration: $-\boldsymbol{\Omega}\times\left(\boldsymbol{\Omega}\times \textbf{r}\right)$. Please note that we obtained that in a purely mathematical and non-intuitive way. We simply manipulated the derivatives. We wish to do better in the next sections. But first, let us multiply with the mass $m$ of the particle and we have:
\begin{equation}
m\textbf{a}=m\textbf{A}-2m\boldsymbol{\Omega}\times\textbf{v}-m\boldsymbol{\Omega}\times\left(\boldsymbol{\Omega}\times \textbf{r}\right).
\end{equation}
The first term on the right is the force $\textbf{F}=m\textbf{A}$ acting on the particle. We finally obtain the second Newton's law in a rotating reference frame:
\begin{equation}
m\textbf{a}=\textbf{F}-2m\boldsymbol{\Omega}\times\textbf{v}-m\boldsymbol{\Omega}\times\left(\boldsymbol{\Omega}\times \textbf{r}\right).
\end{equation}
The second term is the Coriolis force $\textbf{F}_{C}=-2m\boldsymbol{\Omega}\times\textbf{v}$, and the third term is the centrifugal force $\textbf{C}=-m\boldsymbol{\Omega}\times\left(\boldsymbol{\Omega}\times \textbf{r}\right)$. The centrifugal force obviously points away from the axis of rotation. We finally have:
\begin{equation}
m\textbf{a}=\textbf{F}+\textbf{F}_{C}+\textbf{C}.
\end{equation}
This is how we have to correct the second Newton's law in rotating frames of reference - we must add the Coriolis and centrifugal forces to the external force $\textbf{F}$.

\section{Three special scenarios}
Let us have a disc, which rotates around its geometrical center with angular velocity $\boldsymbol{\Omega}$. We show this in FIG. 4.

\begin{figure}[tb]
\includegraphics[width= 0.8\columnwidth]{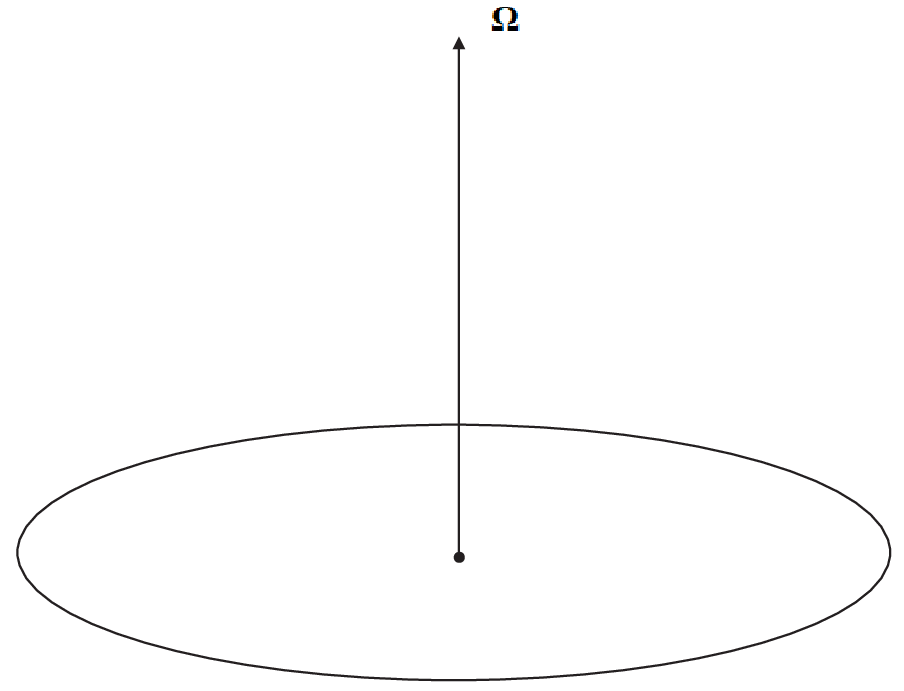}
\caption{A disc rotates around its axis with angular velocity $\boldsymbol{\Omega}$.}
\end{figure}

In this section we shall consider \textit{three} special scenarios. 

\textit{Scenario 1:} let a small ball sits on the disk and \textit{it does not move} according to the \textit{stationary} observer. If we ignore friction between the ball and the disk, we have the following. In the stationary reference frame, the ball does not move. According to the rotating observer, the ball rotates. This means that a \textit{centripetal} acceleration (not centrifugal) should be present in this rotating frame. This centripetal acceleration is what rotates the ball in the rotating frame. Indeed, using equations we have:
\begin{equation}
\textbf{a}=-2\boldsymbol{\Omega}\times\textbf{v}-\boldsymbol{\Omega}\times\left(\boldsymbol{\Omega}\times \textbf{r}\right).\label{a11}
\end{equation}
The first term is the Coriolis acceleration and the second term is the centrifugal acceleration. However, in this Scenario 1, we can calculate $\textbf{v}$. Indeed, according to the rotating observer, the ball rotates in the opposite direction of the disc (see  FIG. 5). At the position of the particle $\textbf{v}_{\text{disk}}=\boldsymbol{\Omega}\times\textbf{r}$ and then:
\begin{equation}
\textbf{v}=-\textbf{v}_{\text{disk}}=-\boldsymbol{\Omega}\times\textbf{r}.\label{v11}
\end{equation}

\begin{figure}[tb]
\includegraphics[width= 0.8\columnwidth]{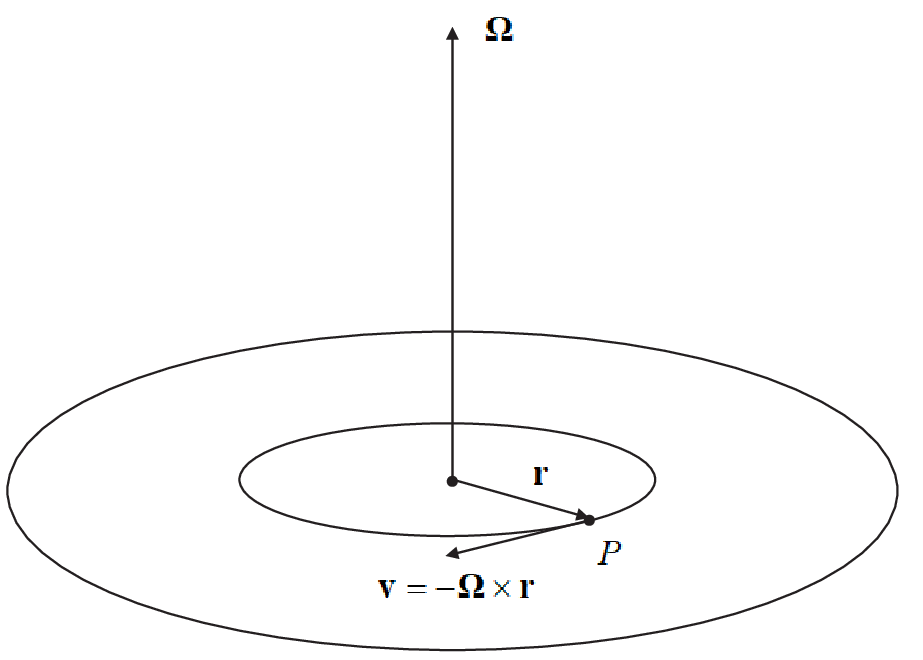}
\caption{Scenario 1: A ball does not move in the stationary frame. According to the rotating frame though, the ball rotates in the opposite direction of the disc with velocity $\textbf{v}=-\boldsymbol{\Omega}\times \textbf{r}$.}
\end{figure}
 
If we substitute Eq. \eqref{v11} into Eq. \eqref{a11} we obtain:
\begin{align}
\textbf{a}=-2\boldsymbol{\Omega}\times\left(-\boldsymbol{\Omega}\times\textbf{r}\right)-\boldsymbol{\Omega}\times\left(\boldsymbol{\Omega}\times \textbf{r}\right)=\boldsymbol{\Omega}\times\left(\boldsymbol{\Omega}\times \textbf{r}\right).
\end{align}

As expected, we have a \textit{centripetal} acceleration, \textit{not} centrifugal. Note that the Coriolis force is a centripetal force and twice larger than the centrifugal force. The net result of these two forces is just a centripetal force.

\textit{Scenario 2:} The ball \textit{does not move} with respect to the \textit{rotating} observer. In  other words, the ball rotates according to the stationary observer. Since, the ball rotates in the stationary reference frame, a centripetal acceleration is present in the stationary frame:
\begin{equation}
\textbf{A}=\boldsymbol{\Omega}\times\left(\boldsymbol{\Omega}\times \textbf{r}\right).
\end{equation}
In addition, since the ball does not move in the rotating frame, we have that $\textbf{v}=0$. In other words, the Coriolis acceleration is $-2\boldsymbol{\Omega}\times \textbf{v}=0$. There is then only a centrifugal acceleration on top of $\textbf{A}$ and we have
\begin{equation}
\textbf{a}=\textbf{A}-\boldsymbol{\Omega}\times\left(\boldsymbol{\Omega}\times \textbf{r}\right)=\boldsymbol{\Omega}\times\left(\boldsymbol{\Omega}\times \textbf{r}\right)-\boldsymbol{\Omega}\times\left(\boldsymbol{\Omega}\times \textbf{r}\right)=0.
\end{equation}
As expected the acceleration $\textbf{a}=0$ in this scenario.

\textit{Scenario 3:} From the point of view of the \textit{rotating} observer the ball moves in radial direction with constant velocity $\textbf{v}=\text{const}$. In other words $\textbf{r}=\textbf{v}t$. This may happen for instance if the ball moves along a small tube, which rotates with the disc and if we place this tube along the radial direction. We show this scenario in FIG. 6.

\begin{figure}[tb]
\includegraphics[width= 0.8\columnwidth]{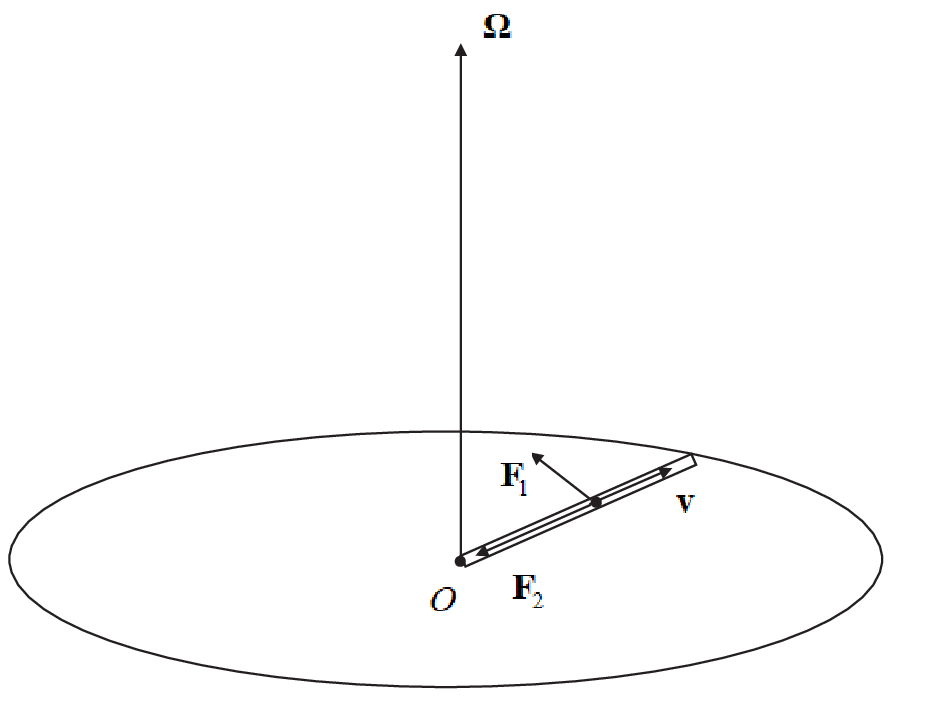}
\caption{A tube rotates with the disc. In the tube a ball moves with constant velocity $\textbf{v}$ (according to the rotating frame). The tube acts on the ball with a force $\textbf{F}=\textbf{F}_{1}+\textbf{F}_{2}$. Here $\textbf{F}_{1}$ is perpendicular to the tube, while $\textbf{F}_{2}$ is alongside the tube. It turns out that $\textbf{F}_{1}$ is equal in magnitude to the Coriolis force. The ball acts back on the tube and might even bend the tube, which is a proof of the presence of the Coriolis force.}
\end{figure}
In that case, it is obvious that $\textbf{v}=\text{const}.$ and $\textbf{a}=0$. However $\textbf{A}\neq 0$. The acceleration $\textbf{A}$ is caused by the force $\textbf{F}=m\textbf{A}$, that the tube impresses upon the ball. In the rotating frame we have
\begin{equation}
0=\textbf{F}+\textbf{F}_{C}+\textbf{C}.
\end{equation}
We can split the force $\textbf{F}=\textbf{F}_{1}+\textbf{F}_{2}$ into two vectors (which we explain  below). Then we have,
\begin{equation}
\textbf{F}_{1}+\textbf{F}_{2}+\textbf{F}_{C}+\textbf{C}=0.
\end{equation}
We wish that $\textbf{F}_{1}$ equalizes the Coriolis force, i.e. $\textbf{F}_{1}+\textbf{F}_{C}=0$ and $\textbf{F}_{2}$ equalizes the centrifugal force, i.e. $\textbf{F}_{2}+\textbf{C}=0$. In that case we have $\textbf{F}_{1}=-\textbf{F}_{C}$ and $\textbf{F}_{2}=-\textbf{C}$. But the Coriolis force $\textbf{F}_{C}$ is perpendicular to the tube, while the centrifugal force $\textbf{C}$ points in the radial direction.  Hence (see FIG. 6) $\textbf{F}_{1}$ should also be \textit{perpendicular} to the tube and $\textbf{F}_{2}$ should be \textit{in the direction} of the tube. The force $\textbf{F}_{1}$ acts on the ball and by the third Newton's principle the ball acts back upon the tube with force $\textbf{P}=-\textbf{F}_{1}$. In other words, the \textit{tube} is pressed by a force $\textbf{P}=-\textbf{F}_{1}=\textbf{F}_{C}$, \textit{perpendicular} to the tube. Here $\textbf{P}$ acts on the \textit{tube}, while $\textbf{F}_{C}$ acts on the \textit{ball}. Then we can detect the presence of the Coriolis force by the presence of the force $\textbf{P}$, which may even slightly bend the tube. This  bending proves the existence of the Coriolis force.
 
\section{Intuitive derivation of the Coriolis and the centrifugal force}

In this section we \textit{derive} \textit{intuitively} the Coriolis acceleration.

For simplicity of notation we shall use \textit{subscript} for the time-dependence. For instance, $\textbf{r}_{t}$ is the position vector at time $t$, while $\textbf{r}_{t+\delta t}$ is the position vector at time $t+\delta t$. We consider small $\delta t$, and at the end of the calculations we take the limit $\delta t\rightarrow 0$.

Now, obviously $\textbf{R}_{t}=\textbf{r}_{t}$ for any moment of time $t$. If we wait a short time interval $\delta t$, we have:
\begin{equation}
\textbf{r}_{t+\delta t}\approx\textbf{r}_{t}+\textbf{V}_{t}\delta t.\label{rt+dt}
\end{equation}
The particle simply moves from point $M$ to point $N$. We can see that in FIG. 7.
\begin{figure}[tb]
\includegraphics[width= 0.65\columnwidth]{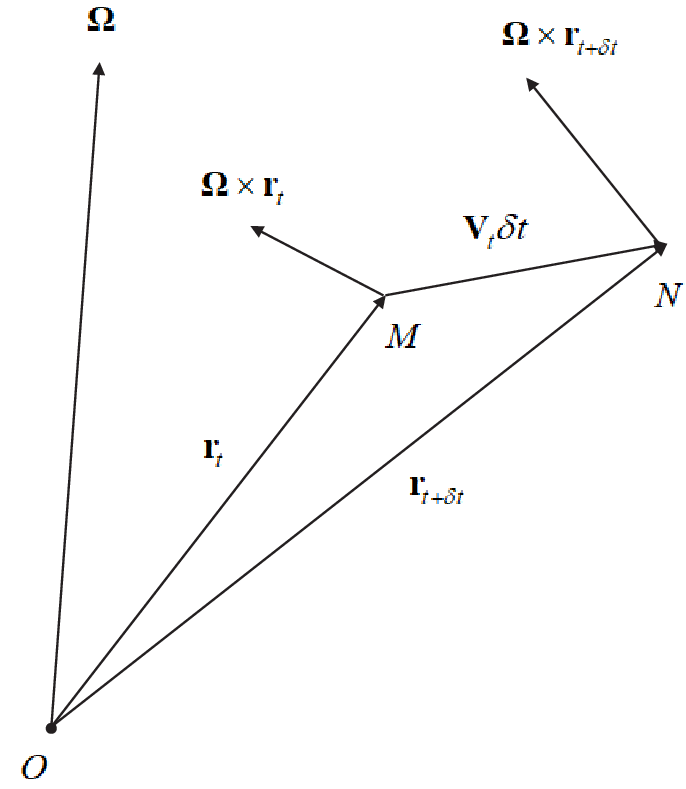}
\caption{A particle moves from point $M$ with a position vector $\textbf{r}_{t}$ to a point $N$ with a position vector $\textbf{r}_{t+\delta t}=\textbf{r}_{t}+\textbf{V}_{t}\delta t$. At point $M$ the local velocity of rotation of the rotating frame is $\boldsymbol{\Omega}\times \textbf{r}_{t}$, while at point $N$ it is different and it is $\boldsymbol{\Omega}\times \textbf{r}_{t+\delta t}$.}
\end{figure}
On the other hand, the velocity $\textbf{V}_{t+\delta t}$ of the particle according to the \textit{stationary} observer changes to:
\begin{equation}
\textbf{V}_{t+\delta t}=\textbf{V}_{t}+\textbf{A}\delta t.\label{Vt+dt}
\end{equation}

So far this is what happens in the stationary frame.  Let us see now what happens in the rotating frame. We claim that:
\begin{equation}
\textbf{v}_{t}=\textbf{V}_{t}-\boldsymbol{\Omega}\times\textbf{r}_{t} \label{vN}
\end{equation}
We shall prove this equation in an intuitive way. 

First, we observe that at the position $\textbf{r}_{t}$, the \textit{reference frame} rotates with velocity $\boldsymbol{\Omega}\times\textbf{r}_{t}$ (see FIG. 7). We shall call this velocity `local rotational velocity`. It is local, since it depends on the particle's position $\textbf{r}_{t}$. In the \textit{new} position of the particle $\textbf{r}_{t+\delta t}$ the local velocity of the rotating frame is \textit{different}: $\boldsymbol{\Omega}\times\textbf{r}_{t+\delta t}$ (see FIG. 7).

Second, according to the rotating observer it is the \textit{particle} that rotates (not the coordinate system) in the \textit{opposite} direction. This will lead to \textit{additional} contribution to the particle's velocity: $-\boldsymbol{\Omega}\times\textbf{r}_{t}$. This contribution is \textit{opposite} to the local rotational velocity of the rotating frame. The \textit{total} velocity according to the rotating observer then becomes: $\textbf{v}_{t}=\textbf{V}_{t}-\boldsymbol{\Omega}\times\textbf{r}_{t}$. This is in fact the same equation as Eq. \eqref{Vv}. Note that this is just the law of addition of velocities in different reference frames. The relative velocity in this case is $\boldsymbol{\Omega}\times\textbf{r}_{t}$ and it is \textit{position-dependent}.

We claim that after a small time interval $\delta t$:
\begin{equation}
\textbf{v}_{t+\delta t}=\textbf{V}_{t+\delta t} -\boldsymbol{\Omega}\times \textbf{r}_{t+\delta t}+\delta\textbf{v}_{1}+\delta\textbf{v}_{2},
\end{equation}
where $\delta\textbf{v}_{1}$ and $\delta\textbf{v}_{2}$ are two small contributions, which we shall calculate below. 

Now, let us focus on the \textit{first} contribution: $-\boldsymbol{\Omega}\times \textbf{r}_{t+\delta t}$. It is due to the fact that the particle has traveled to the new point $N$ (cf. FIG. 7) and it `sees` a different local velocity $\boldsymbol{\Omega}\times \textbf{r}_{t+\delta t}$. If we substitute Eq. \eqref{rt+dt} into this contribution we have:
\begin{equation}
\textbf{v}_{t+\delta t}=\textbf{V}_{t+\delta t}-\boldsymbol{\Omega}\times \textbf{r}_{t}-\boldsymbol{\Omega}\times\textbf{V}_{t}\delta t+\delta \textbf{v}_{1}+\delta \textbf{v}_{2}\label{allcontr}
\end{equation}
The third term of this equation will lead to \textit{half} of the Coriolis acceleration! \textit{This means that half of the Coriolis acceleration is indeed due to the fact that the particle simply travels to a new position and `sees` a new local velocity of rotation of the coordinate system}.

However, there are \textit{other} contributions $\delta\textbf{v}_{1}$ and $\delta \textbf{v}_{2}$, which we shall calculate now.

In FIG. 3 we considered how an arbitrary vector field $\textbf{B}$ rotates when a coordinate system rotates. The vector $\textbf{V}_{t}$ should in the same way be  subjected to purely geometrical rotation to an angle $\delta\boldsymbol{\varphi}=-\boldsymbol{\Omega}\delta t$. There is minus sign in this rotation. This is due to the fact that according to the rotating observer, it is the \textit{particle} that rotates in the \textit{opposite} direction of $\boldsymbol{\Omega}$, \textit{not} the reference frame. This leads to \textit{another} purely geometric contribution:
\begin{equation}
\delta\textbf{v}_{1}=\delta\boldsymbol{\varphi}\times \textbf{V}_{t}=-\boldsymbol{\Omega}\times \textbf{V}_{t}\delta t.
\end{equation}
As we shall see below it is this contribution that will lead to the \textit{second half} of the Coriolis acceleration!

Now, finally we have a \textit{centripetal} acceleration. Indeed, even in the case when $\textbf{V}_{t}=0$ (as in Scenario 1 in the previous Section), a particle will nonetheless rotate with respect to the rotating observer. This will lead to centripetal acceleration and thus to additional velocity after a time $\delta t$:
\begin{equation}
\delta \textbf{v}_{2}=\boldsymbol{\Omega}\times\left(\boldsymbol{\Omega}\times \textbf{r}_{t}\right)\delta t.
\end{equation}

Now, we add \textit{all} contributions in Eq. \eqref{allcontr} and we finally have:
\begin{align}
&\textbf{v}_{t+\delta t}=\textbf{V}_{t+\delta t} -\boldsymbol{\Omega}\times \textbf{r}_{t}-2\boldsymbol{\Omega}\times \textbf{V}_{t}\delta t+\boldsymbol{\Omega}\times\left(\boldsymbol{\Omega}\times \textbf{r}_{t}\right)\delta t\label{Finalv}
\end{align}
The mysterious factor of 2 in the last equation is now finally understood. There are \textit{two} contributions - one due to fact that the particle simply moves to the new position as it travels from point $M$ to point $N$ (cf.  FIG.7) for a time interval $\delta t$. At point $N$, the particle `sees` a new relative velocity $\boldsymbol{\Omega}\times \textbf{r}_{t+\delta t}$. The second contribution is due to mere geometrical rotation of the vector $\textbf{V}_{t}$.  Finally, if we use Eqs. \eqref{vN} and \eqref{Finalv}, we can derive the acceleration $\textbf{a}$ in the rotating frame:
\begin{equation}
\textbf{a}\approx\frac{\textbf{v}_{t+\delta t}-\textbf{v}_{t}}{\delta t}=\frac{\textbf{V}_{t+\delta t}-\textbf{V}_{t}}{\delta t}-2\boldsymbol{\Omega}\times\textbf{V}_{t}+\boldsymbol{\Omega}\times\left(\boldsymbol{\Omega}\times \textbf{r}_{t}\right)
\end{equation}
We take the limit $\delta t\rightarrow 0$ and \textit{remove the subscripts}.
\begin{equation}
\textbf{a}=\textbf{A}-2\boldsymbol{\Omega}\times\textbf{V}+\boldsymbol{\Omega}\times\left(\boldsymbol{\Omega}\times \textbf{r}\right).\label{af1}
\end{equation}
Here we have used the fact that 
\begin{equation}
\textbf{A}=\lim_{\delta t\rightarrow0}\frac{\textbf{V}(t+\delta t)-\textbf{V}(t)}{\delta t}.
\end{equation}

We substitute $\textbf{V}=\textbf{v}+\boldsymbol{\Omega}\times \textbf{r}$ from Eq. \eqref{vN} into Eq. \eqref{af1} and we finally obtain:
\begin{equation}
\textbf{a}=\textbf{A}-2\boldsymbol{\Omega}\times\textbf{v}-\boldsymbol{\Omega}\times\left(\boldsymbol{\Omega}\times \textbf{r}\right),
\end{equation}
which is the same as Eq. \eqref{afinal}. The second term is again the Coriolis acceleration, while the third term is the centrifugal acceleration. Multiplying this equation with the mass $m$ of the  particle will lead to the Coriolis force $\textbf{F}_{C}=-2m\boldsymbol{\Omega}\times\textbf{v}$ and the centrifugal force $\textbf{C}=-m\boldsymbol{\Omega}\times\left(\boldsymbol{\Omega}\times \textbf{r}\right)$. The advantage of this new method of derivation however is that we finally have an \textit{intuitive} understanding of the Coriolis force.

\section{Conclusion}
In this paper we have derived intuitively the Coriolis acceleration $-2\boldsymbol{\Omega}\times\textbf{v}$. We have shown that the factor 2 comes from the reason that there are \textit{two} separate causes that unite into a single combined effect, which is the Coriolis acceleration. The first cause is that the particle travels to a new position, and there it `sees` a different relative velocity. This leads to the first half of the Coriolis acceleration $-\boldsymbol{\Omega}\times\textbf{v}$. Next, while the rotating frame rotates with angular velocity, from the standpoint of the rotating observer, the velocity vector $\textbf{V}$ of the particle simply rotates in the opposite direction and $\delta\boldsymbol{\varphi}\times\textbf{V}=-\boldsymbol{\Omega}\times\textbf{V}\delta t$. It turns out that this leads to another contribution to the acceleration $-\boldsymbol{\Omega}\times\textbf{v}$. Both of these separate contributions unite and the final Coriolis acceleration becomes $-2\boldsymbol{\Omega}\times\textbf{v}$. Not only that but as an additional effect, the centrifugal acceleration has also been derived. This derivation is contrasted with the standard rigorous but not very intuitive derivation of the Coriolis and centrifugal acceleration. We believe that \textit{both} derivations have their merits and \textit{both} might be necessary for better understanding of the fictitious forces in undergraduate courses on Classical Mechanics and Fluid Dynamics. 

\section{Acknowledgment}
This research did not receive any specific grant from funding agencies in the public, commercial, or not-for-profit sectors.

\section{Data Availability Statement}
The author confirm that the data supporting the findings of this study are available within the article and/or its supplementary materials.


\end{document}